\documentclass[conference]{IEEEtran}
\IEEEoverridecommandlockouts
\usepackage{cite}
\usepackage{tabularx,booktabs}
\usepackage{amsmath,amssymb,amsfonts}
\usepackage{algorithmic}
\usepackage{graphicx}
\usepackage{bbding}
\usepackage{textcomp}
\usepackage{multirow}
\usepackage{xcolor}
\usepackage[misc]{ifsym}
\usepackage{geometry}

\begin{document}

\title{Adaptive Learning via a Negative Selection Strategy for Few-Shot Bioacoustic Event Detection\\
 
\author{
    \IEEEauthorblockN{Yaxiong Chen$^{1,2,3}$, Xueping Zhang$^{1}$, Yunfei Zi$^{1}$, $^{*}$Shengwu Xiong$^{2,1,3}$\thanks{* Corresponding Author}}    
    \IEEEauthorblockA{$^1$ School of Computer Science and Artificial Intelligence, Wuhan University of Technology, Wuhan 430070, China}
    \IEEEauthorblockA{$^2$ Sanya Science and Education Innovation Park, Wuhan University of Technology, Sanya 572000, China}
    \IEEEauthorblockA{$^3$ Shanghai Artificial Intelligence Laboratory, Shanghai, 200232, China}
    \IEEEauthorblockA{\{chenyaxiong, xpzhang, yfzi\}@whut.edu.cn,  $^{*}$\{xiongsw\}@whut.edu.cn}
}

\thanks{This work was supported in part by the National Key Research and Development Program of China under Grant 2022ZD0160604; in part by the National Natural Science Foundation of China (NSFC) under Grant 62101393, Grant 62176194, and Grant 62271484; in part by the Project of Sanya Yazhou Bay Science and Technology City under Grant SCKJ-JYRC-2022-17 and Grant SKJC-2022-PTDX-031; in part by the Key Research and Development Program of Hubei Province under Grant 2023BAB083; in part by the Hainan Province “Nanhai New Star” Technology Innovation Talent Platform Project under Grant NHXXRCXM202361; in part by the Youth Fund Project of Hainan Natural Science Foundation under Grant 6220N344; in part by the Knowledge Innovation Program of Wuhan-Basic Research; in part by the Sanya Science and Education Innovation Park of Wuhan University of Technology under Grant 2022KF0020; in part by the High-performance Computing Platform of YZBSTCACC; in part by the Natural Science Foundation of Chongqing under Grant cstc2021jcyjmsxmX1148; in part by the National Science Fund for Distinguished Young Scholars under Grant 61925112; and in part by the Key Research and Development Program of Shaanxi under Grant 2023-YBGY-225.}
}

\maketitle

\begin{abstract}
Although the Prototypical Network (ProtoNet) has demonstrated effectiveness in few-shot biological event detection, two persistent issues remain. Firstly, there is difficulty in constructing a representative negative prototype due to the absence of explicitly annotated negative samples. Secondly, the durations of the target biological vocalisations vary across tasks, making it challenging for the model to consistently yield optimal results across all tasks. To address these issues, we propose a novel adaptive learning framework with an adaptive learning loss to guide classifier updates. Additionally, we propose a negative selection strategy to construct a more representative negative prototype for ProtoNet. All experiments ware performed on the DCASE 2023 TASK5 few-shot bioacoustic event detection dataset. The results show that our proposed method achieves an F-measure of 0.703, an improvement of 12.84\%.
\end{abstract}


\section{Introduction}
Biological Event Detection (BED) aims to locate the start and end time of specific biological vocalisations within a continuous recording. The supervised BED model demands hundreds of annotated examples. However, collecting a sizable training set of biological vocalizations is often impractical due to uneven species abundance and the costly, time-consuming nature of audio annotation \cite{b_baseline}. Consequently, the research on BED with few-shot learning\cite{b_few1,b_few2} is of substantial significance. In the setting of few-shot BED, each evaluation task consists of a single recording, which is further divided into two distinct parts: support set and query set. The support set contains several ($\leq 5$) target annotations while the query set does not contain any annotations. The objective of few-shot BED is to utilize the limited annotations from the support set to locate the start and end time of the target biological vocalisations within the query set.

ProtoNet\cite{b_PrototNet} is a common method for few-shot learning and has proven effective for few-shot BED \cite{b_frame, b_Transductive}. In DCASE 2023 Challenge Task 5 \footnote{https://dcase.community/challenge2023/task-few-shot-bioacoustic-event-detection}, both the official baseline and several submitted solutions \cite{b_Moummad2023, b_XuQianHu2023} also utilized ProtoNet. In ProtoNet, each class is represented by a prototype, which is essentially an average high-dimensional feature vector derived from examples belonging to that class. During inference, the classification results are obtained by calculating the distance between the high-dimensional features of the test data and the corresponding prototypes of different classes. Recently, some flexible learning methods have been applied to few-shot BED to improve ProtoNet. You et al. \cite{b_Transformer} leveraged transfer learning to apply pre-trained models on audio spectrograms to a subset of annotated recordings to classify different events. Yang et al. \cite{b_mutual} proposed a mutual learning framework to continuously update the feature extractor and class prototypes. Wang et al. \cite{b_active} combined few-shot learning with active learning, where the model actively queried human users for annotations of the most informative unannotated data to improve model performance.

Although the above researches have achieved some success in this task, there are still some shortcomings that can be improved. When randomly selecting segments from the query set as negative samples, there is a risk of inadvertently including positive samples, especially in tasks where there is a high proportion of target events. However, if only the background sound segments in the support set are selected as negative samples, the negative prototype obtained may lack representativeness due to an insufficient number of negative samples or variations in background sounds within the query set. Moreover, as a feature extractor, Convolutional Neural Networks (CNN) are good at capturing local features, while pre-trained transformer-based models are good at extracting contextual information from long sequences and tend to produce more confident predictions\cite{b_Transformer}. Therefore, it is challenging to consistently yield optimal results across all tasks with varying durations. Recently, adaptive learning\cite{b_adaptive} has been applied to the field of artificial intelligence with highly competitive results\cite{b_adaptive1,b_adaptive2}. Adaptive learning uses data-driven instruction to adaptively learn experience to meet the individual needs of each student. Inspired by the advantage of adaptive learning, in this study, three novel investigations are proposed to construct a more representative negative prototype and effectively handle tasks with varying durations:
\begin{itemize}
\item We propose a negative selection strategy aimed at augmenting negative samples to construct a more comprehensive negative prototype. 
\item We propose an adaptive learning framework with a teacher model and a student model to update the classifier of ProtoNet, enabling the student model to effectively handle tasks with varying durations.
\item We propose an adaptive learning loss function that adjusts the degree of knowledge transfer from the teacher model based on the durations.
\end{itemize}

\section{Method}

\subsection{Few-shot Setting}

In few-shot learning, we are given a training set and a test set. The training set contains a number of recordings, each given with the full annotations. The test set contains several independent recordings, each representing a test task. Each recording of a test task has only the first few annotations. We refer to the part of the test task with annotations as support set (denoted $S$) and the part without annotations as query set (denoted $Q$).

Let $X$ denote the random variable associated with the acoustic features in $S\cup Q$. $f_{\phi}: X \rightarrow Z \subset \mathbb{R}^{d}$ represents the encoder function, serving as a feature extractor implemented by a deep neural network. In this context, $\phi$ represents the trainable parameters of the network and $Z$ represents the embedded features in $d$-dimensions real number space. $f_{\phi}$ is first trained with the standard cross-entropy loss function by training set.

When dealing with each specific few-shot task, a classifier is initialized, which is parameterized by a weight matrix $W = [w_1, w_2, ..., w_K] \in \mathbb{R}^{K \times d}$. The definition of $w_k$ is as follows:
\begin{eqnarray}
w_k=\dfrac{1}{|S_k|}\sum_{x_i\in S_k}{z_i} , z_i = \frac{f_{\phi}(x_i)}{||f_{\phi}(x_i)||_2} \label{eqw}
\end{eqnarray}
where $x_i$ denotes the acoustic feature of the $i\text{-}th$ example in the $k\text{-}th$ class and $z_i$ represent L2-normalized embedding feature. In this paper, our objective is to determine whether the sound segment is a positive sample. Therefore, we set $K = 2$. 

The prediction probability distribution for a sound segment is computed as: 
\begin{eqnarray}
p_{ik}=\frac{\exp(d(w_k, z_i))}{\sum_{c=1}^K \exp(d(w_c, z_i))}\label{eqq}
\end{eqnarray}
where $d(x, y)$ represents the Euclidean distance between $x$ and $y$. 

\subsection{Adaptive Learning Framework}

\begin{figure*}[ht]
\centerline{\includegraphics[width=1\textwidth]{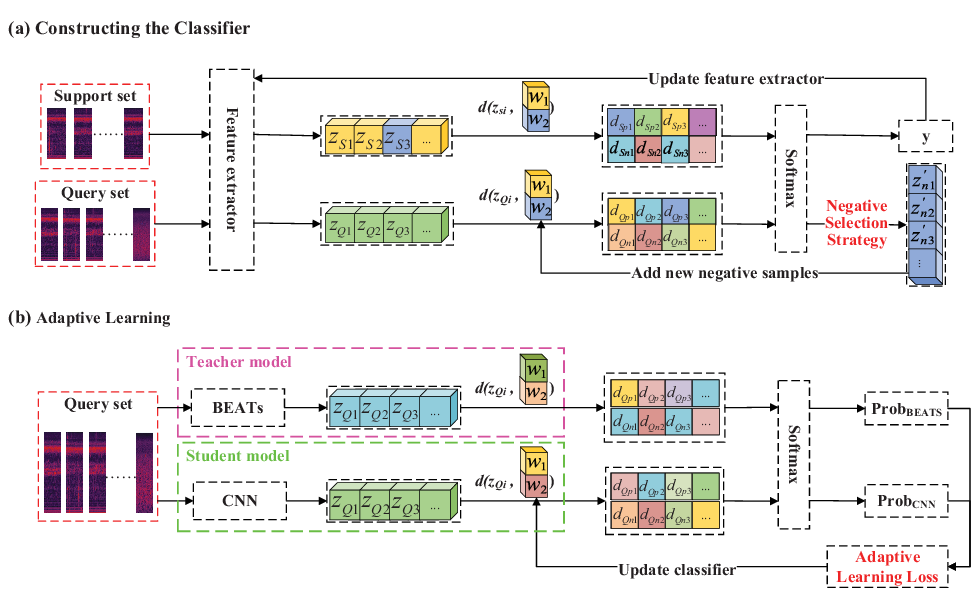}}
\caption{The architecture of proposed adaptive learning framework for few-shot bioacoustic event detection.}
\label{fm}
\end{figure*}

Fig.~\ref{fm} illustrates the process of the proposed adaptive learning framework for bioacoustic event detection. The framework consists of two parts: constructing the classifier and adaptive learning.

\subsubsection{Constructing the Classifier}

For each test task, it first initializes the classifier $W_0$ by the features extracted from the positive and negative samples in support set. Subsequently, a binary classification task is created with the data in support set, and $f_{\phi}$ is updated with cross-entropy loss to better extract features suitable for the specific task. Then a new classifier $W_1$ is constructed with new features extracted from support set by the updated $f_{\phi}$.

\textbf{Negative Selection Strategy} In order to enhance the representativeness of the negative prototype, we propose a negative selection strategy to choose the sound segments $Z_{n}^{\prime}$. The $Z_{n}^{\prime}$ is defined as follows:
\begin{eqnarray}
\begin{aligned}
Z_{n}^{\prime} =&
\{z_j\mid d(z_j,w_{1})>d(w_{1},w_{2}) \cap\\ 
&d(z_j,w_{1})-d(z_j,w_{2})>d(w_{1},w_{2})/2, 0<j\leq |Q|\} \label{eqn}
\end{aligned}
\end{eqnarray}
where $z_j$ is the L2-normalized embedding feature of the $j\text{-}th$ sample in query set, and $w_1$ and $w_2$ represent positive and negative prototypes, respectively. $|Q|$ is the total number of query segments. $d(w_{1},w_{2})$ is the Euclidean distance between positive and negative prototypes.

Let $N$ and $P$ represent the number of negative and positive samples, respectively. $N^{\prime}$ represents the number of new negative samples, and $N^{\prime}$ should satisfy the following condition:
\begin{eqnarray}
 B \leq N^{\prime} \leq P-N+B  \label{eqc}
\end{eqnarray}
where $B$ is a hyper-parameter, in our experiments $B = 5$.

After selecting new negative samples from the query set, they are added to the original negative set and the combined set is used to construct a new classifier $W_2$ for the final test.

\subsubsection{Adaptive Learning} 

Considering that different tasks have different segment lengths, it can be challenging to obtain optimal results on all tasks with varying durations. To improve the model's ability to handle varying duration tasks, we employ adaptive learning to update the classifier. Firstly, both student and teacher models predict the probability that the segments in the query set belong to the positive and negative class. Then, a duration-based adaptive learning loss is computed and used to update the classifier.

\textbf{Teacher and Student Model} The pre-trained Bidirectional Encoder representation from Audio Transformers (BEATs) \cite{b_BEATs} is a transformer-based model designed for audio classification where the acoustic tokenizer and audio Semi-Supervised Learning (SSL) model are optimized through iterations. BEATs is pre-trained on AudioSet\cite{b_AudioSet} by knowledge distillation\cite{b_distilling} and achieves state-of-the-art results on various audio classification benchmarks, even outperforming previous models using more training data and model parameters. We refer to the BEATs model and its associated classifier as ``BEATs-Classifier$_{\mathrm{BEATs}}$", considering it as the teacher model. The backbone network of 4-CNN is the same as \cite{b_PrototNet}, containing only 4 convolutional layers. We refer to the CNN model and its corresponding classifier as ``CNN-Classifier$_{\mathrm{CNN}}$", treating it as the student model.

\textbf{Adaptive Learning Loss} The student model often falls short of expectations when predicting long-duration target events, while the teacher model excels at handling long-duration tasks. Therefore, in cases where the prediction results diverge, we propose a loss function as follows: 
\begin{eqnarray}
Loss=\frac{seg\_len}{T}*(D_{KL}-\lambda*I(X_{st};Y_{st})) \label{eqa}
\end{eqnarray}
In Eq.~\eqref{eqa}, $D_{KL}$ denotes the Kullback-Leible divergence (KL-divergence)\cite{b_kl}, which can be computed as: 
\begin{eqnarray}
D_{KL}=-\sum_{i=1}^{|Q|}\sum_{k=1}^{K}p_{st\_ne\_ik}\log(\frac{p_{te\_ne\_ik}}{p_{st\_ne\_ik}}) \label{eqkl}
\end{eqnarray}
where $p_{st\_ne\_ki}$ represents the predicted probability from the student model, while $p_{te\_ne\_ki}$ represents the predicted probability from the teacher model. $I(X_{st};Y_{st})$ denotes the student model's mutual information entropy\cite{b_I} between the query samples and their latent labels which can be computed as: 
\begin{eqnarray}
I(X_{st};Y_{st})=H(Y_{st})-H(Y_{st}\mid X_{st}) \label{eqi}
\end{eqnarray}
$H(Y_{st})$ is the empirical label-marginal entropy, which is defined as: 
\begin{eqnarray}
H(Y_{st})=-\sum_{k=1}^{K}\hat{p}_{st\_ne\_k}\log(\hat{p}_{st\_ne\_k}) \label{eqhp}
\end{eqnarray}
where $\hat{p}_{st\_ne\_k}$ is marginal distribution, defined as:
\begin{eqnarray}
\hat{p}_{st\_ne\_k}=\frac{1}{|Q|}\sum_{i=1}^{|Q|}p_{st\_ne\_ik} \label{eqp}
\end{eqnarray}
$H(Y_{st}\mid X_{st})$ is an empirical estimate of the conditional entropy of labels given the query acoustic features, which can be calculated as: 
\begin{eqnarray}
H(Y_{st}\mid X_{st})=-\frac{1}{|Q|}\sum_{i=1}^{|Q|}\sum_{k=1}^{K}p_{st\_ne\_ik}\log(p_{st\_ne\_ik}) \textbf{\label{eqh}}
\end{eqnarray}

 In our experiments, $\lambda$ is set to 0.5 and ${T}$ is a constant set to 150. Only if there is a deviation prediction between the teacher model and the student model, the above loss is calculated and the model parameters are updated by gradient descent. In this way, the teacher model and the student model serve as mutual supervisors, even in the situation where there are no labels. Furthermore, only update the parameters of Classifier$_{\mathrm{CNN}}$, while all other parameters remain frozen. KL-divergence is used to assess the dissimilarity of the predicted probability distribution between the student model and teacher model. The adaptive weight parameter $\frac{seg\_len}{T}$ is used to adjust the degree of knowledge transfer from the teacher model to the student model. $I(X_{st};Y_{st})$ quantifies the level of correlation between the query samples and their latent labels. Greater correlation implies greater confidence in the model's predictions, and this measure is applied to bolster the confidence of the student model.

\section{Experiments and Results}

\subsection{Experimental Setups}

\textbf{Dataset} The dataset used in our study is from DCASE 2023 Task 5 challenge\footnote{https://dcase.community/challenge2023/task-few-shot-bioacoustic-event-detection\#development-set} which is a few-shot bioacoustic event detection dataset including a training set and a test set. The training set consists of 174 recordings across various bioacoustic vocalisations. The test set consists of 18 recordings with three distinct subsets: ME, PB, and HB. Event durations vary among different recordings, with median event durations of 0.14-0.24ms for ME, 0.02-0.17ms for PB, and 1.3-19.35ms for HB. Each recording has additional annotations, including the start and end time of sound events, as well as event names. The training set is fully annotated, whereas the test set provides annotations for only the first five positive instances.

\textbf{Processing} All recordings are down-sampled to a 16kHz sampling rate and subsequently transformed into 128-dimensional Mel-Spectrograms. This transformation is accomplished by applying the Short-Time Fourier Transform (STFT) with a frame shift of 10ms and a frame length of 25ms. Following this Mel-Spectrogram generation, the Per-Channel Energy Normalization (PCEN) is extracted. 

\textbf{Dynamic Segment Setup} In the test set, the durations of the target events vary across different tasks. Using excessively long segment length may lead to mismatches, especially when dealing with short-duration events. Conversely, if the segment length is too short, valuable contextual information may be lost. To address this issue, we set dynamic segment lengths for different test recordings. For each test recording, we get a medium event duration on the first five positive events and then set final segment and hop length for that recording, according to Table~\ref{tabs_seg_len}. $m$ is the median duration of the first five positive events. $seg\_len$ is the segment length. Event median duration is counted in number of frames.

\begin{table}[htbp]
\caption{Dynamic segment and hop length on FSL tasks.}
\setlength{\tabcolsep}{2pt}
\begin{center}
\resizebox{\linewidth}{!}{
\begin{tabular}{|c|c|c|c|c|c|}
\hline
\textbf{Median duration} & 0-20     & 20-100   & 100-200   & 200-400   & \textgreater400 \\ \hline
\textbf{Segment length}  & 20       & $m$      &  $m/2$    & $m/4$     & $m/8$           \\ \hline
\textbf{Hop length}      & \multicolumn{5}{c|}{$seg\_len$ / 4}                 \\ \hline
\end{tabular}
}
\label{tabs_seg_len}
\end{center}
\end{table}

\textbf{Training} We train the feature extractor directly with cross-entropy loss. To explain further, the feature extractor is followed by a dense layer, and a softmax layer is added to get classification probabilities. For the CNN model, the segment length of the training data is 20 frames, while for the BEATs model, the segment length of the training data is 128 frames. We utilize the Adam optimizer with an initial learning rate of $1 \times 10^{-4}$ for training and $1 \times 10^{-5}$ for updating the feature extractor and classifier.

\textbf{Metrics} In test tasks, after obtaining positive probability predictions for test segments, segments with a probability lower than 0.5 are filtered out. Subsequently, adjacent segments are merged into events. As in the DCASE challenge guideline\footnote{https://dcase.community/challenge2023/task-few-shot-bioacoustic-event-detection\#evaluation-metric}, an Intersection Over Union (IOU) value of 0.3 is used to define event matching. For all experiments, the evaluation metric employed is the event-level F-measure\cite{b_f1}.

\subsection{Experimental Results}

\begin{table}[htbp]
\caption{F-measure comparison of different methods.}
\setlength{\tabcolsep}{3pt}
\begin{center}
\resizebox{\linewidth}{!}{
\begin{tabular}{|c|c|c|c|c|}
\hline
\multicolumn{1}{c|}{\textbf{Method}}                           & \multicolumn{1}{c}{\textbf{ME}} & \multicolumn{1}{c}{\textbf{PB}} & \multicolumn{1}{c}{\textbf{HB}} & \multicolumn{1}{c}{\textbf{ALL}}        \\ \hline
\multicolumn{1}{c|}{Baseline\cite{b_baseline}}                 & \multicolumn{1}{c}{0.613} & \multicolumn{1}{c}{0.289} & \multicolumn{1}{c}{0.196}    & \multicolumn{1}{c}{0.295}           \\
\multicolumn{1}{c|}{TIM+SpecAugment\cite{b_banoori2024few}}    & \multicolumn{1}{c}{0.415} & \multicolumn{1}{c}{0.392} & \multicolumn{1}{c}{0.720}    & \multicolumn{1}{c}{0.560}           \\
\multicolumn{1}{c|}{BED-DFSL\cite{b_Continual} }               & \multicolumn{1}{c}{0.829} & \multicolumn{1}{c}{0.408} & \multicolumn{1}{c}{0.771}    & \multicolumn{1}{c}{0.606}           \\ 
\multicolumn{1}{c|}{\textbf{NSS+AL (Ours)}}                             & \multicolumn{1}{c}{0.916} & \multicolumn{1}{c}{0.511} & \multicolumn{1}{c}{0.821}    & \multicolumn{1}{c}{\textbf{0.703}}  \\ \hline
\end{tabular}
}
\label{tabs_res}
\end{center}
\end{table}

\begin{figure}[htbp]
\centerline{\includegraphics[width=0.5\textwidth]{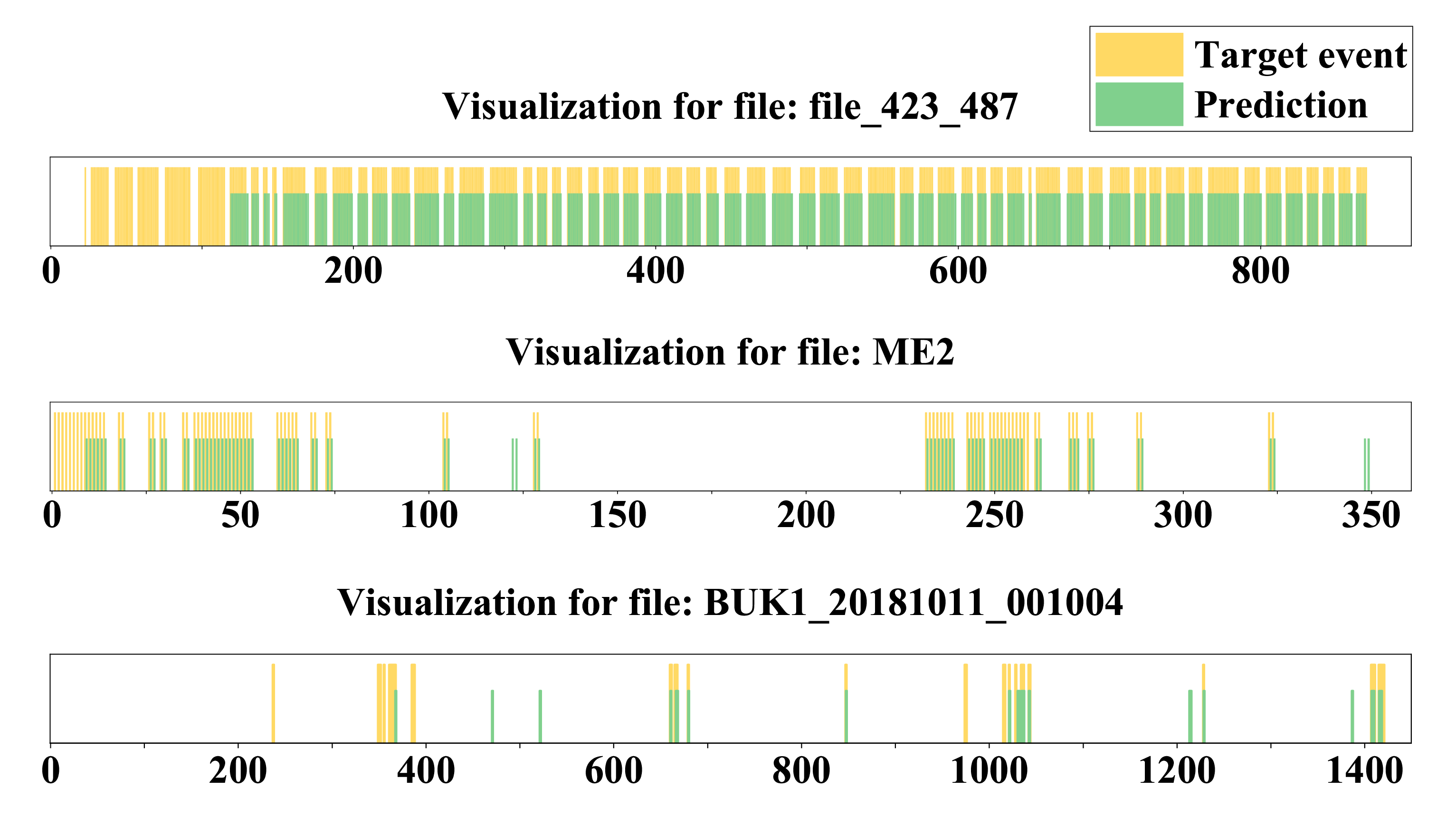}}
\caption{Model prediction performance visualisation.}
\label{fp}
\end{figure}

Table~\ref{tabs_res} presents the experimental results of different methods. Our approach achieves a 0.703 F-measure on the test set of DCASE 2023 TASK5 dataset, significantly outperforming the baseline\cite{b_baseline}. The baseline used a meta-learning strategy to train a 4-layer CNN model as its feature extractor and randomly selected sound segments as negative samples from the query set. Compared with the latest methods, the F-measure of our method is 25.5\% better than TIM+SpecAugment\cite{b_banoori2024few} and 16.0\% better than BED-DFSL\cite{b_Continual}. TIM+SpecAugment used a transudative inference with data augmentation. BED-DFSL modified Dynamic Few-Shot Learning (DFSL) to generalize it to the BED task with weight alignment loss and positive enhancement. The result is remarkable and shows that our proposed method achieves better performance compared to other methods, highlighting its effectiveness.

To visualize the effectiveness of our proposed adaptive learning framework for few-shot bioacoustic event detection, we illustrate the comparison between the positive events predicted by our method and the true target events, as shown in Fig.~\ref{fp}. Task file\_423\_487, ME2 and BUK1\_20181011\_001004 are from the subsets of HB, ME and PB, respectively. The first five events are from the support set, so there are no predictions on them. The predicted positive events are the results of amalgamating the predicted positive segments. As can be seen, our framework demonstrates high detection accuracy, bringing us closer to identifying target events. Task file\_423\_487 exhibits longer durations and a higher density of target events, which explains the reduced effectiveness of the baseline within the HB subset, because a substantial portion of randomly selected sound segments from the query set are positive samples.

\subsection{Ablation Study}

\begin{table}[htbp]
\caption{Ablation study results for negative selection Strategy and adaptive learning.}
\begin{center}
\resizebox{\linewidth}{!}{
\begin{tabular}{c|cccc}
\hline
\textbf{Methods} & \textbf{ME} & \textbf{PB} & \textbf{HB} & \textbf{ALL} \\ \hline
None             & 0.893       & 0.407       & 0.327       & 0.453            \\
NS               & 0.886       & 0.423       & 0.757       & 0.623            \\
NSS              & 0.913       & 0.431       & 0.816       & 0.647            \\ \hline
AL (ONLY)        & 0.916       & 0.421       & 0.812       & 0.638            \\
NSS+AL           & 0.916       & 0.511       & 0.821       & 0.703            \\ \hline
\end{tabular}
}
\end{center}
\label{tabs_ablation}
\end{table}

Table~\ref{tabs_ablation} presents the results of the ablation study. We separately examined the effects of the negative selection strategy and adaptive learning on the results. Ultimately, we demonstrate that each component of our method contributes positively to the overall improvement of the results.

\textbf{Negative selection strategy} We first utilize a 4-layer CNN model as PortoNet’s feature extractor which is trained with the cross-entropy loss function. The test data set use dynamic segment length. After constructing the classifier, the feature extractor is updated with a binary classification task. The sound segments used to construct the negative prototype are randomly selected from the query set. The F-measure obtained through this approach is 0.453. NS utilizes background sound segments from the support set as negative samples to construct the negative prototype. NSS extends NS by incorporating the negative selection strategy to augment negative samples from the query set. Ultimately, they achieve F-measure of 0.623 and 0.647, respectively. Our proposed negative selection strategy improves F-measure by 3.85\% on this task. This suggests that the negative sample selection strategy can select appropriate negative samples, help to construct a more representative negative prototype and better perform classification tasks.

\begin{table}[htbp]
\caption{F-measure comparison of teacher model, student model and adaptive learning results.}
\setlength{\tabcolsep}{3pt}
\begin{center}
\resizebox{\linewidth}{!}{
\begin{tabular}{c|c|cc|cccc}
\hline
\multirow{2}{*}{\textbf{\begin{tabular}[c]{@{}c@{}}Exp.\\ No.\end{tabular}}} & \multirow{2}{*}{\textbf{\begin{tabular}[c]{@{}c@{}}Feature \\ extractor\end{tabular}}} & \multicolumn{2}{c|}{\textbf{Methods}}        & \multicolumn{4}{c}{\textbf{F-measure}}                \\ \cline{3-8} 
                                   &                                 & \textbf{NSS}    & \textbf{AL}      & \textbf{ME}  & \textbf{PB}  & \textbf{HB}  & \textbf{ALL}  \\ \hline
1                                  & BEATs                           & \XSolidBrush     & \text{-}         & 0.537        & 0.203        & 0.831        & 0.376             \\
2                                  & CNN                             & \XSolidBrush     & \XSolidBrush     & 0.886        & 0.423        & 0.757        & 0.623             \\
3                                  & CNN                             & \XSolidBrush     & \CheckmarkBold   & 0.916        & 0.421        & 0.812        & 0.638             \\ \hline
4                                  & BEATs                           & \CheckmarkBold   & \text{-}         & 0.496        & 0.304        & 0.835        & 0.461             \\
5                                  & CNN                             & \CheckmarkBold   & \XSolidBrush     & 0.913        & 0.431        & 0.816        & 0.647             \\
6                                  & CNN                             & \CheckmarkBold   & \CheckmarkBold   & 0.916        & 0.511        & 0.821        & 0.703             \\ \hline
\end{tabular}
}
\label{tabs_comp}
\end{center}
\end{table}

\textbf{Adaptive Learning} AL (ONLY) indicates that only the adaptive learning method is used, which ultimately achieves an F-measure of 0.625, an improvement of 2.41\% compared to NS. NSS+AL indicates the combination of the negative selection strategy and adaptive learning, which results in an F-measure of 0.703, an overall improvement of 12.84\%. 

Table~\ref{tabs_comp} presents the F-measure comparison of teacher model, student model and adaptive learning results. In Exp.No. 1, 2 and 3, do not use the negative selection strategy, while in Exp.No. 4, 5 and 6, do use the negative selection strategy. Although the overall performance of the teacher model with BEATs could not match the student model with CNNs before adaptive learning, the teacher model exhibits superior performance on the HB subset which has longer target event durations. After adaptive learning, the student model improves the F-measure not only on the HB subset but also on the ME and PB subsets. This improvement persists even when the teacher model underperforms on ME and PB subsets, possibly due to mutual supervision between the two models. The results suggest that adaptive learning enhances the model's ability to handle tasks with varying durations.

\section{Conclusion}
In this paper, we propose a novel adaptive learning framework to improve ProtoNet for few-shot BED. To construct a more representative negative prototype, we propose a negative selection strategy to select more representative negative samples. To address the challenge posed by tasks with varying durations of target events, we employ adaptive learning and propose a duration-based adaptive learning loss to adjust the extent of knowledge transfer from the teacher model during updating the classifier. Our method ultimately achieves an F-measure of 0.703 on the DCASE 2023 TASK5 dataset.

\bibliographystyle{IEEEtran}
\bibliography{IEEEref}
\vspace{12pt}
\end{document}